# Effects of non-equilibrium in ultrafast irradiation of matter

Nikita Medvedev*[a,b]

[a] Institute of Physics, Czech Academy of Sciences, Na Slovance 1999/2, 18221 Prague 8, Czechia;

[b] Institute of Plasma Physics, Czech Academy of Sciences, Za Slovankou 3, 18200 Prague 8, Czechia;

## ABSTRACT

This proceeding discusses nonequilibrium effects in matter exposed to XUV/X-ray irradiation. When ultrashort, intense XUV/X-ray pulses interact with materials, they trigger a complex sequence of processes, including electronic excitation, nonequilibrium electron kinetics, energy exchange with the atomic system, electronic thermalization, and subsequent atomic dynamics. These effects were investigated using XTANT-3, a hybrid simulation tool that simultaneously models all relevant dynamics. XTANT-3 integrates (a) a Monte Carlo transport method for photon absorption and fast electron kinetics, (b) the Boltzmann equation for nonequilibrium slow electron dynamics, (c) a transferable tight-binding approach for electronic structure evolution and interatomic potential modeling, and (d) molecular dynamics for atomic system response. This approach enables a detailed study of nonequilibrium effects in each subsystem and their interplay with nonthermal damage, where electronic excitation alters the interatomic potential. Methods of quantifying the nonequilibrium in the electronic and atomic subsystems are discussed.

## 1. INTRODUCTION

Intense ultrashort XUV and X-ray pulses are produced at modern free-electron laser (FEL) facilities, with the typical parameters of photon energies from a few tens of eV up to few tens of keV; pulse duration ~10 fs; and intensities sufficient to induce phase transitions in a single shot (and even turn the material into hot plasma) [1–4]. Using FEL pulses in a controlled way enables exciting and probing transient nonequilibrium states of matter. Such a precise and powerful tool for studies of nonequilibrium kinetics and producing of unusual material states found practical applications, as well as advanced our understanding of fundamental physics [5–10]. Materials treatment with conventional (optical and infrared) laser irradiation is commonly employed for processing, creating nano-technology, medical applications, and many others [8,11–13]. In turn, ultrafast X-ray irradiation advanced science towards femtosecond and angstrom precision measurements of both atomic and electronic kinetics [1–3].

An ultrashort pulse of XUV/X-ray photons triggers complex processes in target material. Photons are absorbed by core atomic shells, creating core holes and emitting fast electrons [10,14]. The photoelectrons then undergo a cascading process of impact ionization of secondary electrons [15]. This process creates new holes in the core shells and the valence band of the solid. Simultaneously, the quasi-elastic scattering of electrons transfers energy to the atomic system – a process known as electron-ion or electron-phonon coupling [5,16]. Core-shell holes decay *via* the Auger channel, emitting new free electrons, or fluorescence, emitting photons; for the typical FEL parameters, the Auger-decay channel is dominant [17,18].

Electronic coupling to the atomic system increases the kinetic energy of atoms, which may eventually overcome the barriers of melting or solid-solid phase transition and induce observable changes in the material. At the same time, the excitation of electrons modifies the interatomic potential, which – if sufficiently strong – may also lead to material damage via bond breaking and atomic rearrangement. This process is known as nonthermal melting, or, more generally, nonthermal phase transition [16,19–22].

* Corresponding author: ORCID: 0000-0003-0491-1090, email: nikita.medvedev@fzu.cz

Excited electrons transiently depart from their equilibrium state – they acquire non-Fermi-Dirac distribution, which then relaxes towards the equilibrium one over time. After an FEL irradiation, typically, the shape of the distribution function may be described as bump-on-hot-tail: slow electrons are close to equilibrium, but the fraction of fast electrons is strongly nonequilibrium [23,24]. Yet, the effect of non-equilibrium in the slow electron fraction may also be pronounced at certain conditions – for example, in bandgap materials, where the electronic relaxation between the valence and the conduction band electrons may require a long time.

Such excited electrons are out of equilibrium with the atomic system – the fact that forms the foundation of the widely-used two-temperature model [25,26]. However, the atomic system itself may also be out of equilibrium [27]. Thus, after irradiation, three kinds of nonequilibrium are transiently present: nonequilibrium within the electronic system, nonequilibrium between electrons and atoms, and nonequilibrium within the atomic system. An interplay of all these effects creates unusual kinetics, which may result in the formation of unusual states of matter.

Below, we discuss our recent works on simulation of the nonequilibrium effects in matter under FEL irradiation. We discuss how the nonequilibrium can be quantified in various simulation methods and the consequences of the nonequilibrium states on observable effects [27,28].

## 2. MODEL AND THEORY

To study the effects of nonequilibrium in material response to FEL irradiation, the hybrid code XTANT-3 is used [29,30]. The code and physical models in it are thoroughly described in its manual; see Ref. [30]. Let us briefly recall the relevant models used, as they define the methods needed for assessing and quantifying the nonequilibrium.

The code combines a few separate methods on the fly:

(i) Transport Monte Carlo (MC) scheme simulates the photoabsorption, secondary electronic cascades, and Auger decays of core holes [29,30]. Fast electrons are traced until their energy falls below the curoff energy, typically ~10 eV, counted from the bottom of the conduction band (or the Fermi level for metals). When an electron loses its energy below the threshold, it joins the low-energy fraction of electrons. The scattering of such electrons on low-energy electrons drives the low-energy fraction of electrons out of equilibrium, delivering energy to them.

(ii) The Boltzmann equation (BE) is used to simulate the kinetics of low-energy electrons, populating the valence and the conduction band [28]. It includes the electron-electron scattering via the relaxation-time approximation [28], the electron-phonon (electron-ion) coupling[28] *via* the so-called dynamical coupling approach [31], and the source terms calculated from the MC module above.

(iii) The transferable tight binding (TB) method is applied to evaluate the transient electronic structure and the interatomic forces [30,32]. The electronic structure, or the molecular orbitals, defined as the eigenstates of the transient Hamiltonian, depend on the positions of all the atoms in the simulation box and thus evolve in time with the atomic motion. The interatomic forces depend on the electronic populations on these energy levels and thus respond to changes in both the energy levels and the electron distribution function (traced with BE above), which enable the modeling of nonthermal phase transitions [29].

(iv) The molecular dynamics (MD) method is employed to propagate the atomic trajectories. The interatomic potential or forces acting on atoms are calculated with the help of the abovementioned TB method. The energy transfer from electrons *via* elastic scattering of high-energy electrons in the MC and electron-ion coupling in the BE is delivered to atoms at each timestep *via* the velocity scaling [29].

We emphasize that both subsystems, the electronic and the atomic one, are modelled with methods capable of treating nonequilibrium conditions – they do not assume thermal equilibrium in the system. However, since the electrons and atoms are treated with different appropriate methods, they require different descriptions of the nonequilibrium.

Electrons are treated at the level of their distribution function (in BE), and thus it allows us to directly evaluate the electronic entropy, $S_e$, on each timestep of the simulation:

$$S_e = -k_B 2 \sum [(f_e/2) \ln(f_e/2) + (1 - f_e/2) \ln(1 - f_e/2)] \qquad (1)$$

here $k_B$ Is the Boltzmann constant, $f_e$ is the electron distribution function, and factors of 2 are due to the normalization of the distribution function according to the spin degeneracy [28].

The maximum entropy corresponds to the thermal equilibrium. Thus, the transient entropy may be compared to the maximal possible electron entropy at the corresponding timestep, defined by using the equivalent Fermi-Dirac function in Eq.(1). To construct the equivalent equilibrium distribution, the equivalent temperature ($T_e$) and chemical potential (µ) are defined from the first and second moments of the distribution function:

$$\begin{cases} n_e = \sum f_e(\varepsilon_i, t) = \sum f_{eq}(\varepsilon_i, \mu, T_e, t) \\ E_e = \sum \varepsilon_i f_e(\varepsilon_i, t) = \sum \varepsilon_i f_{eq}(\varepsilon_i, \mu, T_e, t) \end{cases} \quad (2)$$

Where for the given total number of electrons in the system ($n_e$) and their energy ($E_e$), the Eqs.(2) are solved numerically; $\varepsilon_i = \langle i|H_{TB}|i\rangle$ are the electronic energy levels, the eigenfunctions of the TB Hamiltonian [28,32].

Table 1. Various temperatures in a two-component system (atoms and electrons, Eq.(3)) and a single-component system (only atoms, Eq.(4)).

| Vector field $B(\vec{R}, \vec{P})$ | Title | One-component definition, Eq.(4) | Two-component definition, Eq.(3) |
|---|---|---|---|
| $\vec{P}$ | Kinetic temperature | $T_{kin} = \frac{2}{3}\langle E_{kin}\rangle$ | $T_{kin} = \frac{2}{3}\langle E_{kin}\rangle$ |
| $P^2\vec{P}$ | Fluctuational temperature | $T_{fluc} = \sqrt{2/3}\sqrt{\langle E_{kin}^2\rangle - \langle E_{kin}\rangle^2}$ | $T_{fluc} = \sqrt{2/3}\sqrt{\langle E_{kin}^2\rangle - \langle E_{kin}\rangle^2}$ |
| $\boldsymbol{\nabla}_{\vec{R}}H_{tot}$ | Configurational temperature | $T_{config}^{(1)} = \frac{\langle(\boldsymbol{\nabla}_{\vec{R}}H)^2\rangle}{\langle\boldsymbol{\nabla}_{\vec{R}}^2H\rangle}$ | $T_{config} = -\frac{\langle\vec{F}\cdot[\vec{F}_i + \frac{1}{2}\vec{F}_{ie}]\rangle}{\langle\boldsymbol{\nabla}_{\vec{R}}\cdot\vec{F}\rangle + \frac{1}{2T_e}\langle\vec{F}\cdot\vec{F}_{ie}\rangle}$ |
| $F^2\vec{F}$ | Hyperconfigurational temperature | $T_{hyperconf}^{(1)} = -\frac{\langle F^4\rangle}{\langle\boldsymbol{\nabla}_{\vec{R}}\cdot(F^2\vec{F})\rangle}$ | $T_{hyperconf} = -\frac{\langle(F^2\vec{F})\cdot[\vec{F}_i + \frac{1}{2}\vec{F}_{ie}]\rangle}{\langle\boldsymbol{\nabla}_{\vec{R}}\cdot(F^2\vec{F})\rangle + \frac{1}{2T_e}\langle(F^2\vec{F})\cdot\vec{F}_{ie}\rangle}$ |
| $\vec{R}$ | Virial temperature | $T_{vir}^{(1)} = -\frac{1}{3}\langle\vec{R}\cdot\vec{F}\rangle$ | $T_{vir} = -\frac{\langle\vec{R}\cdot[\vec{F}_i + \frac{1}{2}\vec{F}_{ie}]\rangle}{3 + \frac{1}{2T_e}\langle\vec{R}\cdot\vec{F}_{ie}\rangle}$ |

\* Some rearrangement and combination of the terms with $B(\vec{R},\vec{P}) = \vec{P}$ required to construct the fluctuational temperature.

$E_{kin}$ is the kinetic energy of an atom; $\vec{F}$ is the total force acting on the atom; $\vec{F}_{ie}$ is only the electronic part of the force. Note that the virial temperature definition does not apply to periodic boundary conditions in MD simulations [27].

The atoms are treated as strongly interacting individual particles in MD; the calculation of their single-particle distribution function is impossible. Instead, alternative methods are used to quantify their degree of nonequilibrium [27]. The nonequilibrium in a particle-based simulation may be assessed *via* a comparison of various nonequilibrium temperatures – they differ in nonequilibrium states, but all coincide, merging into a single thermodynamic temperature, in the equilibrium [33]. We recently extended the definition of such a generalized temperature from classical MD simulations to the system of atoms and electrons [27]:

$$T_i = \frac{\langle B(\vec{R},\vec{P})\cdot\boldsymbol{\nabla}_{\vec{R}}[U_i+\frac{1}{2}V_{ei}]\rangle}{\langle\boldsymbol{\nabla}_{\vec{R}}\cdot B(\vec{R},\vec{P})\rangle-\frac{1}{2T_e}\langle B(\vec{R},\vec{P})\cdot\boldsymbol{\nabla}_{\vec{R}}V_{ei}\rangle} \quad (3)$$

Where $T_i$ is the generalized atomic temperature; $\boldsymbol{B}(\vec{R},\vec{P})$ is an arbitrary vector field in the phase-space; $U_i$ is the potential energy of interaction between the atoms; $V_{ei}$ is the potential energy of interaction between the atoms and electrons, and the angle brackets $\langle\ldots\rangle$ denote an ensemble average [27].

As shown in Ref.[27], this definition (3) reduces to the standard single-component definitions (only atoms, no electrons, for classical MD) in the case of $T_e \equiv T_i$:

$$T_i^{(1)} = \frac{\langle \boldsymbol{B}(\vec{R},\vec{P}) \cdot \nabla_{\vec{R}} H_{tot}\rangle}{\langle \nabla_{\vec{R}} \cdot \boldsymbol{B}(\vec{R},\vec{P})\rangle} \tag{4}$$

Now, choosing various vector fields $\boldsymbol{B}(\vec{R},\vec{P})$ allows one to obtain an infinite amount of temperature definitions; in particular, all the standard definitions are also restored (see example in Table 1).

## 3. RESULTS

As an example, we will consider diamond irradiated with an FEL pulse with 30 eV photons, 15 fs FWHM duration, and 1.5 eV/atom of the absorbed dose. The tight binding parameters used are identical to the previous works [29,32]. A microcanonical ensemble with 64 atoms in the simulation box is used for this illustration. Martyna-Tuckerman 4th order algorithm for propagation of the atomic trajectories is applied with the timestep of 0.5 fs. The electron-electron relaxation is traced with the relaxation time approximation, using τ=500 fs for a complete (interband) thermalization, with additional separate relaxation of the electrons within the conduction band and the valence band, each with τ=10 fs [28]. The simulation starts at 200 fs before the pulse arrival (counted from the center of the Gaussian pulse). The results are shown in Figure 1.

The electronic temperature rises quickly to the values of $T_e$~24,000 K due to the photoabsorption, see Figure 1(a). Then, the entire system reacts to it. The electrons are coupling to the atoms (phonons), exchanging their kinetic energy, with the rather high coupling parameter, see Figure 1(c).

At the same time, the nonthermal effects start driving the atomic system into the new phase – graphite-like planes (for a detailed description of the nonthermal solid-solid phase transition in diamond, see e.g. [29]). This is accompanied by the bandgap collapse, Figure 1(e). The collapse of the gap turns the originally insulating material, diamond, into a semimetallic phase, graphite. It takes place via both the lowering of the bottom of the conduction band and the simultaneous rise of the top of the valence band, which then meet in the middle, see Figure 1(f). The nonthermal phase transition also provides atoms with additional kinetic energy – the so-called nonthermal acceleration of atoms (for details, see [34]). Both effects, electron-phonon coupling and nonthermal acceleration, lead to an extremely fast increase of the atomic kinetic energy, and, correspondingly, its kinetic temperature (see Figure 1(a,b)).

During this ultrafast phase transition, electrons are out of equilibrium, as can be seen by their entropy behavior in Figure 1(d). The transient entropy is lower than the equilibrium one. The electrons equilibrate by the time of ~50 fs, which is, in a large part, a result of the bandgap collapse: after the collapse and merger of the valence and the conduction bands, it is easy for the electrons to thermalize to a common equilibrium distribution.

The atoms are also driven out of equilibrium by the electronic excitation, see Figure 1(b). Various atomic generalized temperatures start to diverge after the FEL pulse arrival, inducing electronic excitation. The divergence becomes most pronounced during the phase transition itself [27]. This indicates that the phase transition is inherently a nonequilibrium process. Nonetheless, it is interesting to note that the temperatures defined in different subspaces of the phase space partially equilibrate much faster than the complete equilibration takes place: the kinetic and fluctuational temperatures (both defined solely in the momentum space) are close to each other almost at all times; the configurational and hyperconfigurational temperatures (both defined in the configurational, or coordinate, space) are close to each other, but different from the momentum-space temperatures [27]. This finding, reported in Ref. [27], suggests that a nonequilibrium system may be described with just two parameters – generalized temperatures in momentum and configurational subspaces – enabling construction of multi-temperature thermodynamics, which is significantly simpler than a fully nonequilibrium kinetic theory.

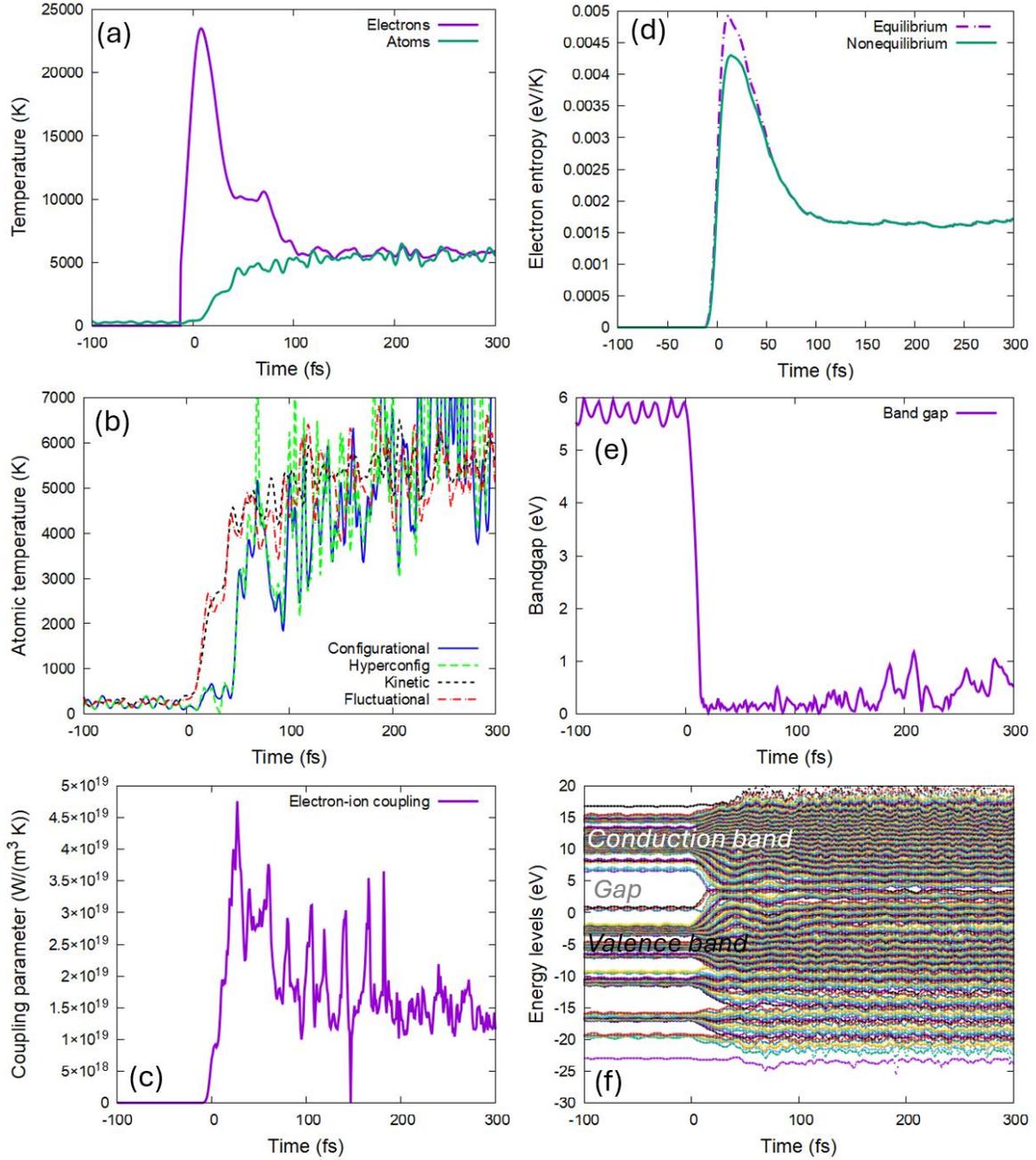

Figure 1. XTANT-3 simulation of diamond irradiated with 15 fs FWHM pulse, 30 eV photon energy, 1.5 eV/atom absorbed dose. (a) Electronic and kinetic atomic temperatures. (b) Kinetic, fluctuational, configurational, and hyperconfigurational atomic temperatures (see Table 1). (c) Electron-ion (electron-phonon) coupling parameter. (d) Transient electronic entropy, and the maximal one corresponding to the equivalent equilibrium Fermi-Dirac distribution. (e) Electronic band gap. (f) Electronic energy levels (molecular orbitals; band structure at the Γ-point).

## 4. CONCLUSIONS

It is discussed that matter under ultrafast XUV/X-ray irradiation transiently enters a nonequilibrium state. In it, each subsystem of the material is out of equilibrium: electrons depart from the equilibrium Fermi-Dirac distribution, atoms cannot be described with a single thermodynamic temperature, and there is a nonequilibrium between the two systems as well [27,28]. The electronic nonequilibrium can be quantified by evaluating the electronic entropy in the methods having access to the electron distribution function [28]. In the methods tracing atoms as individual particles (such as molecular dynamics simulations), the atomic nonequilibrium state may be assessed by applying the generalized temperatures definitions [27]. It is discussed that the atomic system with excited electrons, undergoing nonthermal phase transition, may be adequately described with just two generalized temperatures: in the momentum and configurational subspaces of the phase space [27].

## 5. ACKNOWLEDGMENTS

The financial support from the Czech Ministry of Education, Youth, and Sports (grants No. LTT17015, LM2018114, and No. EF16_013/0001552) is gratefully acknowledged. Computational resources were provided by the e-INFRA CZ project (ID:90254), supported by the Ministry of Education, Youth and Sports of the Czech Republic.